\documentstyle[11pt]{article}
\def\fnote#1#2{\begingroup\def\thefootnote{#1}\footnote{#2}\addtocounter
{footnote}{-1}\endgroup}

\def\gev{{\rm GeV}\,}

\begin{document}

\hfill{UTTG-25-2020}

\begin{center}

{\bf On the Development of Effective Field Theory}

\vspace{20pt}

Steven Weinberg\fnote{*}{Electronic address:
weinberg@physics.utexas.edu}\\
{\em Theory Group, Department of Physics, University of
Texas\\
Austin, TX, 78712}

\vspace{20pt}

{\bf Abstract}

\end{center}


 This is a lightly edited version of the talk given on September 30, 2020 to inaugurate the international seminar series  {\it All Things EFT}.  It reviews some of the early history of effective field theories, and concludes with a discussion of the implications of effective field theory for future discovery.\\

\vspace{15pt}



What is the world made of?  This question  is  perhaps is the deepest and  earliest in all of science.  Greeks were asking this question a hundred years before the time of Socrates. By the time that I became a graduate student an answer had  apparently been settled.  The world is made not of water, earth, air or fire, but of fields. There is the electromagnetic field that when quantum mechanics is applied to it is manifested in the form of bundles of energy, momentum -- particles that are called photons. There is an electron field that similarly when quantized appears as particles called electrons. And there are other fields that we in the late 1950s knew we did not yet know about. The weak and the strong interactions were pretty mysterious. It was clear that there had to be more than just electrons and photons. But we looked forward to a description of nature as consisting fundamentally of fields as the constituents of everything. 

The quantum field theory of electrons and photons in the late 1940s had scored a tremendous success. Theorists -- Feynman, Schwinger, Tomonaga, Dyson -- had figured out after decades of effort how to do  calculations preserving not only Lorentz invariance but also the appearance of Lorentz invariance at every stage of the calculation. This allowed them to sort out the infinities in the theory that had been noticed in the early 1930s by Oppenheimer and  Waller, and that had been the {\it  b\^ete noire} of theoretical physics throughout the 1930s. They were able to show in the late 1940s that these infinities could all be absorbed into a redefinition, called a renormalization, of the electron mass and charge and the scales of the various fields. And they were able to do calculations of unprecedented precision, which turned out to be verified by experiment: calculations of the Lamb shift and the anomalous magnetic moment of the electron. 

More than that, and this particularly appealed to me as a graduate student, renormalization theory didn't always work. It wouldn't work unless the theory had a certain kind of simplicity. 
Essentially  the only coupling constants allowed in the theory, in units where Planck's constant and the speed of light were 1, had to be dimensionless, like the charge of the electron: $e^2/4\pi$ is $1/137$. And this provided not only a means of dealing with the infinities but a rationale for the simplicity of the theory. Of course we always like simple theories. But when we have discovered successful simple theories we always should ask why are they so simple? Renormalizability provided an answer to that question. Of course it was only an answer if we thought that these were really the fundamental theories that described nature at all scales. Otherwise, anything else might intervene to get rid of the infinities.

We hoped that we would see the rest of physics -- the mysterious strong and weak nuclear forces -- brought into a similar framework. And that is indeed what happened in the following decades in the development of the Standard Model. Once we got past the obscurities produced by spontaneous symmetry breaking in the weak interactions and color trapping in the strong interactions, the Standard Model was revealed to us as a theory that was really not very different from quantum electrodynamics. We had more gauge fields, not just the electromagnetic field but gluon and  $W$ and $Z$ fields. There were more fermions, not just the electron but a whole host of charged leptons and neutrinos and quarks. But the Standard Model seemed to be quantum electrodynamics writ large. One could perhaps  have been forgiven for reaching a stage of satisfaction that, although not everything was answered, although there were still outstanding questions, that this was going to be part of nature at the most fundamental level. 

Now that has changed. In the decades since the completion of the Standard Model a new and cooler view has become widespread. The Standard Model, we now see -- we being, let me say, me and a lot of other people -- as a low-energy approximation to a fundamental theory about which we know very little. And low energy means energies much less than some extremely high energy scale $10^{15}-10^{18}\;\gev$. As a low energy approximation we expect corrections to the Standard Model.  These corrections are beginning to show up. Some of them have already been found. 

This whole point of view goes by the name of effective field theory. It's had applications outside elementary particle physics in areas like superconductivity. I am not going to in this talk try to bring the subject up-to-date, including all the applications of effective field theory to hadronic physics, and to areas of physics outside particle physics, like superconductivity. That's going to be done by subsequent lecturers in this series by physicists who played a leading role in the development of effective field theory beyond anything that I knew about it in the early days. They are true experts in the field. I won't dare to try to anticipate what they will say. I'll talk about a subject on which I am undoubtedly the world's expert and that is my own history of how I came to think about these things. I'm a little bit unhappy that I am putting myself too much forward. Other people came to effective field theories through different routes. I'm not going to survey anyone else's intellectual history except my own.

From my point of view, it started with current algebra. The late 1950s and early 1960s were a time of despair about the future -- about the practical application of quantum field theory to the strong interactions. Although we could believe that quantum field theory  was at the root of things we didn't know how to apply quantum field theory to the strong interactions. Perturbation theory didn't work. Instead, a method was developed called current algebra in which one concentrated on the currents of the weak interactions, the vector and axial vector currents, using their commutation relations, their conservation properties and in particular a suggestion made by Nambu that the divergence of the axial vector current was dominated by one pion states. This current algebra was used in a very clunky way, requiring very detailed calculations that just called out for a simpler approach to derive useful results, including the Goldberger-Treiman formula for the pion decay amplitude and  the Adler-Weisberger sum rule  for the axial vector coupling constant. 

After a while some of us began to think that although these results were important and valuable, perhaps we were giving too much attention to the currents themselves, which of course play a central role in the weak interactions. We ought to concentrate on the symmetry properties of the strong interactions which made all this possible. In particular, the existence of a symmetry which gradually emerged in our thinking, chiral $SU(2)\times SU(2)$, which is just isotopic spin symmetry applied separately to the left handed and right handed parts of what we would now say are quark fields. 

This symmetry is a property of the strong interactions which would be important even if there weren't any weak interactions. And it was employed to derive purely strong interaction results, like for example, the scattering lengths of pions on nucleons and pions on pions
and more complicated things like the emission of any number of soft pions in high energy collisions of   nucleons or other particles. When this was done using these, as I said, clunky methods of current algebra, looking at the results, they seemed to look like the results of a field theory. You could write down Feynman diagrams just out of the blue which would reproduce the results of current algebra. 

And so the question naturally arose, is there a way of avoiding the machinery of current algebra by just writing down a field theory that would automatically produce the same results with much greater ease and perhaps physical clarity? Because after all in using current algebra one had to always wave one's hands and make assumptions about the smoothness of matrix elements, whereas if you could get these results from Feynman diagrams you could see what the singularity structure of the matrix elements was and make only those smoothness assumptions that were consistent with that. 

At the beginning this was done using a standard theory with the chiral symmetry that we thought was at the bottom of all these results, the linear sigma model, and then re-defining the fields in such a way that the results would look like current algebra.  The effect of the redefinition was the introduction of a non-linearly-realized chiral symmetry.  Eventually the linear sigma model was scrapped;  instead the procedure was simply to ask, what kind of symmetry transformation for the pion field alone, some transformation into a nonlinear function of the pion field, would have the algebraic properties of  chiral symmetry, based on the Lie algebra of $SU(2)\times SU(2)$. 

That theory had the property that in lowest order in the coupling constant $1/F_\pi$ the results reproduced the results of current algebra.  Why did it? Well, it had to, because a theory having  chiral and Lorentz invariance and  unitarity and smoothness satisfied the assumptions of current algegra and therefore had to reproduce the same results.  For example  current algebra calculations gave a pion-pion scattering  matrix element of order $1/F_\pi^2$, where $F_\pi$ is the pion decay amplitude, so if you use this field theory and just threw away everything except the leading term which is of order  $1/F_\pi^2$, this matrix element had to agree with the results of current algebra. 

In this way phenomenological Lagrangians were developed that could be thought of as merely labor-saving devices, which were guaranteed to give the same results as current algebras because they satisfied the same underlying conditions of symmetry and so on, and that could be used in lowest order because current algebra said the result was of lowest order in  the $1/F_\pi$ coupling -- that is, $1/F_\pi^2$ for $\pi\pi$ scattering and also for pion-nucleon scattering. If you had more pions you would have more powers of $1/F_\pi$. But the results would always agree with current algebra.

No one took these theories seriously  as true quantum field theories at the time. (I am talking about the late 1960s.)  No one would have dreamed at this point of using the phenomenological Lagrangian in calculating loop diagrams. What would be the point? We knew that it was the tree approximation that reproduced current algebra. As I said, these phenomenological Lagrangians were simply labor-saving devices. 

The late 1960s and 1970s saw many of us engaged in the development of the Standard Model. During this time I wasn't thinking much about current algebra or phenomenological Lagrangians. The soft pion theorems had been successful,  not only in agreeing with experiment, but also in killing off a competitor of quantum field theory known as S-matrix theory. S-matrix theory had been the slogan of a school of theoretical physicists headed by Geoff Chew at Berkeley. I had been there at Berkeley in its heyday but had never bought on to it. 

Their idea was that field theory is hopeless. It deals with things we will never observe like quantum fields. What we should do is just to study things that are observable, like  S-matrix elements: apply principles of Lorentz invariance, analyticity and so on, and get results like dispersion relations that we can  compare with observation.  It was even hoped that stable or unstable composite  particles like the $\rho$ meson  provide the force that produces these composites, so that using this bootstrap mechanism one could actually do calculations.  This never really worked as a calculational scheme, but was extremely attractive philosophically because it made do with very little except the most fundamental assumptions, without introducing things like strongly interacting fields that we really didn't know about. But the chiral symmetry results, the soft-pion results, showed that some of the approximations assumed in using S-matrix theory, like  strong  pion-pion interactions at low energy, just weren't right.  Chiral symmetry provided actual calculations of processes like $\pi\pi$ and $\pi$-nucleon scattering, which made the ideas of S-matrix theory seem unnecessary, attractive as the philosophy was.

S-matrix theory had been largely killed off and chiral symmetry had been put in the books as a success, but we were all involved in applying the ideas of quantum field theory in the weak and the electromagnetic interactions and then the strong interactions, building up the Standard Model, which was beginning to be very successful experimentally. It was a very happy time for the interaction between theory and experiment. During this period of course I was teaching.  It was in the course of teaching that my point of view changed, because in teaching quantum field theory  I had to keep confronting the question of the motivation for this theory. Why should these students take seriously the assumptions we were making, in particular the formalism of writing fields in terms of creation and annihilation operators, with their  commutation relations. Where did this come from?  The standard approach was to take a field theory like Maxwell's theory and  quantize it, using the rules of canonical quantization.  Lo and behold, you turn the crank, and out come the commutation relations for the operator coefficients of the  wave functions in the quantum field.

I found that hard to sell, especially to myself. Why should you apply the canonical formalism to these fields? The answer  that the canonical formalism  had proved useful in celestial mechanics in the 19th century wasn't really very satisfying.  In particular, suppose there was a theory that in other ways was successful but couldn't be formulated in terms of canonical quantization, would that bother us? In fact, we have quantum field theories like that. They're not realistic theories. They're theories in six dimensions, or theories we derive by compactifying six dimensional theories. We have quantum field theories that apparently can't be given a Lagrangian formalism -- that can't be derived using the canonical formalism. So I looked for some other way of teaching the subject.

I fastened on a point of view that is really not that different from S-matrix theory. One starts  of course by assuming the rules of quantum mechanics as laid down in the 1920s, together with special relativity and then one makes an additional assumption,  the cluster decomposition principle, whose importance was emphasized to me by a colleague at Berkeley, Eyvind Wichmann, while I was there in the 1960s. The cluster decomposition principle says essentially that unless you make special efforts to produce an entangled situation the results of distant experiments are uncorrelated. The results of an experiment at CERN are not affected by the results being obtained by an experiment being done at the same time at Fermilab.  
The natural way of implementing the cluster decomposition principle is by writing the Hamiltonian as a sum of products of creation and annihilation operators with non-singular coefficients. Indeed, this had been done for many years by condensed matter physicists not because they were interested in quantum field theory as a fundamental principle, but in order to sort out the volume dependence of various thermodynamic quantities.  They were managing this  by introducing creation and annihilation operators long before I began to teach courses in quantum field theory. 

It gradually appeared to me in teaching the subject that although individual quantum field theories like quantum electrodynamcis certainly have content  quantum field theory in itself has no content except the principles on which it's based, namely quantum mechanics, Lorentz invariance, and the cluster decomposition principle, together with whatever other symmetry principles you may want to invoke, like chiral symmetry and gauge invariance.

This means that if we think we know what the degrees of freedom are -- the particles we have to study -- like for example, low energy pions, that if we write down the most general possible theory involving fields for these particles, including all possible interactions consistent with the symmetries, which  in this case are Lorentz invariance and  chiral $SU(2)\times SU(2)$, if we write down all possible invariant terms in the Lagrangian, and we work to all orders in perturbation theory, the result can't be wrong, because it's just a way of implementing these principles. It is just giving you the most general possible matrix element consistent with Lorentz invariance, chiral symmetry, quantum mechanics, cluster decomposition, and  unitarity.

Now, this may not sound as if it's a very useful realization.  If you tell someone to calculate using the most general possible Lagrangian with an unlimited number of free parameters and calculate to all orders in perturbation theory they're likely to seek some advice elsewhere on how to spend their time. But it's not that bad, because even if the theory has no small dimensionless couplings, you can use this approach to generate a power series in powers of the energy that you're interested in. For instance, if you're interested in low-energy pions and you don't want to consider energies high enough so that  $\pi\pi$ collisions can produce nucleon-antinucleon pairs, you will be dealing with typical energies $E$  well below the nucleon mass. This very general Lagrangian gives you a power series in powers of $E$.  Specifically, aside from a term that depends only on the nature of the process being considered, the number of powers of $E$ arising from a given Feynman diagram is the total number of derivatives acting at all the vertices, plus half the number of nucleon lines connected to all the vertices, plus twice the number of loops.  

Now, chiral symmetry dictates that the number of derivatives plus half the number of nucleon fields in each interaction is always equal to or greater than two. The diagrams that give the lowest total number of powers of $E$ are those constructed only from  interactions where that number equals two, with no loops.  For pion-pion scattering the leading term comes from a single vertex with just two derivatives.  For pion-nucleon scattering there is a diagram with  a single vertex with one deerivative to which are connected two nucleon lines.  Thsee diagrams  are the ones that we had  been using since the mid 1960s to reproduce the results of current algebra.

But now the new thing was that you could consider contributions of higher order in energy. If you look for terms that have two additional powers of energy they could come from diagrams where you have one interaction with the number of derivatives plus half the number of nucleon fields equaling not two but four, plus any number of interactions with this number equal to two, and no loops. 
Or you could have only interactions with the numbers of derivatives plus half the number of nucleon fields equal to two, that is, just the basic interactions that reproduce current algebra, plus one loop.  The infinity in the one loop diagram could be canceled by that one additional vertex that has, say, not two derivatives but four derivatives. In every order of perturbation theory as you encounter more and more loops, because you are allowing more and more powers of energy, you get more and more infinities, but there are always counterterms available to cancel the infinities. A non-renormalizable theory, like the soft pion theory, is just as renormalizable as a renormalizable theory.  You have an infinite number of terms in the Lagrangian, but only a finite number are needed to calculate S-matrix elements to any  given order in $E$.

Similar remarks apply to gravitation, which I think has led to a new perspective on general relativity.   Why in the world should anyone take seriously Einstein's original theory, with just the Einstein-Hilbert action in which  only two derivatives act on metric fields? Surely that's just the lowest order term in an infinite series of terms with more and more derivatives.  In such a theory,   loops are made finite by counterterms provided by the higher order terms in the Lagrangian. This   point of view   has been actively pursued by Donoghue and his collaborators. 

Teaching came to my aid again.  At the blackboard one day in 1990 it suddenly occurred to me that one of the kinds of interaction that has the number of derivatives plus half the number of nucleon fields equal to two is the interaction with no derivatives at all and four nucleon fields. It had taken me a decade to realize that four divided by two is two. This sort of interaction  is just the kind of hard-core nucleon-nucleon interaction that nuclear physicists had always known would be needed to understand nuclear forces. But now we had a rationale for it. In a calculation of nuclear forces as a power series in energy, the leading terms are just the ones that the nuclear physicists had always been using, pion exchange plus a hard core.  This point of view has been explored by Ordo\~{n}ez and van Kolck and others.

In the theories that I have been discussing, the chiral symmetry theory of soft pions and general relativity, the symmetries don't allow a purely renormalizable interaction. In a theory of gauge fields and quarks and leptons and scalars you can have a renormalizable Lagrangian.  Throwing away everything but renormalizable interactions you have just the Standard Model, a renormalizable theory of quarks and leptons and gauge fields and a scalar doublet. But now we can look at the Standard Model as one term  in a much more general theory, with all kinds of non-renormalizable interactions, which yield corrections if  higher order in energy.  

In these corrections the typical energy $E$ must be divided by some characteristic  mass scale $M$. For the chiral theory of soft pions this mass scale is  $M\approx 2\pi F_\pi$, about $1200\,{\rm MeV}$. For the theory of weak, strong and electromagnetic interactions $M$ is probably a much higher scale, perhaps something like the scale of order $10^{15}\gev$ where Georgi, Quinn and I found the effective gauge couplings of the weak, strong, and electromagnetic interactions  all come together. Or perhaps it's the characteristic scale of gravitation, the Planck scale $10^{18}\;\gev$.  Or perhaps  somewhere in that general neighborhood.  It is the large scale of $M$ that made it a good strategy in constucting the Standard Model to look for a renaormalizable theory .

We now expect that there are corrections to the renormalizable Standard Model of the order of powers of $E/M$. How in the world are we ever going to find corrections that are suppressed by such incredibly tiny fractions? The one hope  is that those corrections can violate symmetries that we once thought were inviolable, but that we now understand are simply accidents, arising from the constraint of renormalizability that we imposed on the Standard Model.  

Indeed, quite apart from the development of effective field theory, one of the great things about the 
Standard Model was that it explained  various symmetries that could not be fundamental, becasue  we already knew they were only partial or approximate symmetries.  This included  flavor conservation, such as strangeness conservation,  a symmetry of the strong and electromagnetic interactions that was manifestly violated in the weak interactions.  Another example was charge conjugation invariance -- likewise a good symmetry of strong and electromagnetic but violated by weak interactions. The same was true of parity, although in this case you have to make special accommodations for non-perturbative effects.  All thse were accidental symmetries.  imposed  by the simplicity of the Standard Model necessary for renormalizability plus  other symmetries like gauge symmetries and Lorentz invariance that seem truly exact and fundamental.  
Chiral symmetry itself is such an accidental symmetry, though only approximate. It becomes an accidental exact symmetry of the strong interactions in the limit in which  the up and down quark masses are zero, as does isotopic spin symmetry. Since these masses  are not zero but relatively small chiral symmetry is an approximate accidental symmetry.  

Now, coming back to effective field theory, there are other symmetries within the Standard Model that are accidental symmetries of the whole renormalizable theory of weak, strong, and electromagnetic interactions: in particular baryon conservation and lepton conservation are respected aside from very small non-perturbative effects  (well, very small at least in laboratories, though maybe not so small cosmologically). If baryon and lepton conservation  are only accidental properties of the Standard Model maybe they are not symmetries of nature. In this case there is no reason why baryon and lepton conservation should be respected by  non-renormalizable corrections  to the Lagrangian, and so you would expect terms of ${\cal O}(E/M)$ or ${\cal O}((E/M)^2)$ or  higher order as corrections to the Standard Model that violate these symmetries.

Wilczek and Zee and I independently did a catalog of the leading terms of this type. Some of them -- those involving baryon number non-conservation -- give you corrections of ${\cal O}((E/M)^2)$. They have not been yet been discovered experimentally. But there are other terms that produce corrections of ${\cal O}(E/M)$ that violate lepton conservation, and they apparently have been discovered, in the form of neutrino masses. I wish I could say that the effective field theory point of view had predicted the neutrino masses. Unfortunately it must be admmitted that neutrino masses were already proposed by Pontecorvo as a solution to the apparent deficit of neutrinos coming from the Sun. So though I can't say that the effective field theory approach had predicted neutrino masses,  I do think  that it gets a strong boost from the fact that we now  have evidence of a correction to the renormalizable part of the Standard Model. But  of course there already was a known correction.  
Gravitation was always there, warning us that renormalizable quantum field theory can't be the whole story. 

I expect that sooner or later we will be seeing another departure from the renormalizable Standard Model in the discovery of proton decay, or some other example of  baryon nonconservation. In a sense baryon nonconservation has already been discovered, because we know from the present ratio of baryon number to photon number that in the early universe before the temperature dropped to a few GeV there was about 1 excess quark over every $10^9$ quark-antiquark pairs.  This has to be explained by the non-renormalizable corrections to the Standard Model, and indeed it has been explained, unfortunately not just by one such model but by many different models.  We don't know the actual mechanism for producing baryon number in the early universe, but I have no doubt that it will be found.

 There are still unnatural elements in the Standard Model. I said that you would expect the leading terms that describe physical reactions to be given by the renormalizable theory -- here the Standard Model -- with the effects  of  non-renormalizable terms suppressed by powers of $E/M$. Those corrections come from interactions that have coupling constants whose dimensions have negative powers of mass, $1/M$, $1/M^2$, and so on. But what about interactions that have positive powers of mass? Why aren't they there at ${\cal O}(M^n)$? Well unfortunately we don't have a good explanation. 

The cosmological constant is such a term. It has the dimensions of energy per volume, in other words $M^4$. We don't know why it's as small as it is.  There is also the bare mass of the Higgs boson. That's the one  term in the Standard Model Lagrangian whose  coefficient has the dimensionality of a positive power of mass. We don't know why it is not ${\cal O}(10^{15}\gev)$. These are  great mysteries that confront us: Why are the terms in our present theory that have the dimensions of positive powers of mass so small compared to the scale that we think is the fundamental scale, somewhere in the neighborhood of $10^{15}-10^{18}\;\gev$? We don't know. 

With the new approach to the Standard Model I think we have to say that this theory in its original form is not what we thought it was. It's not a fundamental theory. But at the same time I want to stress that the Standard Model will survive in any future textbooks of physics in the same way that Newtonian mechanics has survived as  a theory we use all the time applied to the solar system. All of our successful theories survive as approximations to a future theory. 

There's a school of philosophy of science associated in particular with the name of Thomas Kuhn, that sees the development of science -- particularly of physics -- as a series of paradigm shifts in which our point of view changes so radically that we can barely understand the theories of earlier times. I don't believe it for a minute.  I think our successful theories always survive, as Newtonian mechanics has, and as I'm sure the Standard Model will survive, as approximations. 

Now we have to face the question, approximations to what?  We think the Standard Model is a low energy approximations to a theory whose constituents involve mass scales way beyond what we can approach in the laboratory, scales on the order of $10^{15}$, $10^{18}\gev$. It may be a field theory. It may be an asymptomatically safe field theory, which avoids coupling constants running off to infinity as the energy increases. Or it seems to me more likely that it's not a field theory at all, that it's something like a string theory.  In this case we will understand the very successful field theories with which we work as effective field theories, embodying the principles of quantum mechanics and symmetry, applied as an approximation valid at low energy  where any theory will look like a quantum field theory. 

\vspace{10pt}

\vspace{10pt}

\begin{center}
{\bf 	ACKNOWLEDGMENTS}
\end{center}

This work  is supported by the National Science Foundation under grant number
Phy-1914679 and also with support from the Robert A. Welch Foundation, Grant No. F-0014. 

\end{document}